# Field Effect Transistors on Rubrene Single Crystals with Parylene Gate Insulator

V. Podzorov[1], V. M. Pudalov[1,2], and M. E. Gershenson[1]
[1] *Department of Physics and Astronomy, Rutgers University, Piscataway, New Jersey 08854*
[2] *P. N. Lebedev Physics Institute, 119991 Moscow, Russia*

We report on fabrication and characterization of the organic field effect transistors (OFETs) on the surface of single crystals of rubrene. The *parylene* polymer film has been used as the gate insulator. At room temperature, these OFETs exhibit the p-type conductivity with the field effect mobility up to 1 cm$^2$/Vs and the on/off ratio ~ 10$^4$. The temperature dependence of the mobility is discussed.

An increase of the carrier mobility in the organic field effect transistors (OFETs) is an outstanding problem (for a review, see e.g., Ref. [1, 2, 3]). Solving this problem might open new opportunities for both physics of two-dimensional systems and applications. Relatively high values of the surface carrier mobility (the so-called field-effect mobility) at the room temperature, $\mu(300\,K) \sim 1$ cm$^2$/Vs, have been reported for the OFETs on the basis of highly ordered vacuum-deposited organic films[4]. These values are comparable with the mobility $\mu(300\,K)$ for the optically generated carriers in the volume of organic molecular single crystals, measured by the time-of-flight method[5]. However, at low temperatures, the field-effect mobility in OFETs differs significantly from the mobility of "bulk" carriers in organic single crystals. Indeed, the mobility of the bulk carriers increases with cooling and can be as large as several hundred cm$^2$/Vs at $T < 100$ K [5]. In contrast, the field-effect mobility for the thin-film OFETs, being limited by a large concentration of defects in the vacuum-deposited organic films, usually decreases with cooling[4, 6]. The dependence $\mu(T)$ in the best pentacene thin-film OFETs varies from thermally activated to almost temperature-independent with $\mu(300K)$ increasing from 0.3 cm$^2$/Vs to 1.2 cm$^2$/Vs [4].

It has been recognized that ordering of the organic film structure is very important for achieving high field-effect mobilities[3, 5]. Fabrication of the OFETs at the surface of *single crystals* of high-purity organic compounds might increase the mobility significantly. However, this technological process poses a challenge. In particular, the proper choice of the gate insulator, which would form a trap-free interface with organic crystals, is crucial.

In this Letter, we report on fabrication and characterization of OFETs on the surface of single crystals of rubrene. Use of thin *parylene* films as the gate insulator allowed fabrication of the OFETs with reproducible characteristics. At room temperature, the devices demonstrate transistor characteristics with $\mu$ up to ~ 1 cm$^2$/Vs and the on/off ratios up to 10$^4$. The field-effect mobility has been studied over the temperature range $T = 77 - 300$ K. The observed slow decrease of $\mu$ with cooling is similar to the $\mu(T)$ dependences reported previously for the best organic thin-film FETs [4].

We have used physical vapor transport in hydrogen[7] for the growth of single crystals of rubrene. The purity of the crystals has been tested by measuring the bulk carrier transport prior to the FET fabrication[8]. The 300 Å-thick source and drain contacts were formed at the crystal surface by thermal evaporation of silver through a shadow mask. In order to minimize the heat (radiation) load on the crystal surface, the distance between the thermal evaporation boat and the sample has been increased up to 70 cm, the temperature of the crystal during the deposition was maintained below 0 °C, and the deposition rate did not exceed 2 Å/sec. In addition to the thermally evaporated contacts, we have also used contacts painted on the crystal surface with the aqueous solution of colloidal graphite.

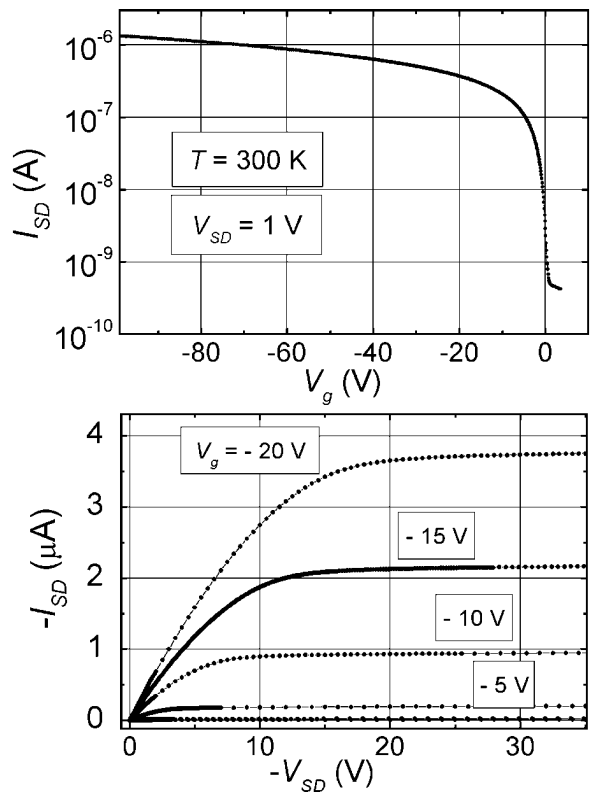

**FIG. 1.** The room temperature characteristics of the OFET fabricated on the surface of rubrene single crystal. The source and drain electrodes are made of colloidal graphite (channel dimensions: $L \approx 0.5$ mm, $W \approx 1$ mm). The upper panel shows the source-drain current, $I_{SD}$, versus the gate

voltage, $V_g$. The lower panel: $I_{SD}$ versus the bias voltage $V_{SD}$ at several fixed values of $V_g$.

The polymer *parylene* has been chosen as the gate insulator material. This material forms a pinhole-free conformal coating as thin as 0.1 μm with excellent dielectric and mechanical properties [9]. Parylene was deposited onto the rubrene crystals with preformed source and drain contacts in a home-made reactor. The crystals were kept at room temperature during the deposition. The parylene thickness ~ 0.2 μm was sufficient to cover uniformly even the rough colloidal-graphite contacts. The output of working devices with the parylene gate insulator approached 100%. The parylene films deposited onto organic crystals withstand multiple temperature cycling between 300 K and 4.2 K. As the final step of FET fabrication, a 500 Å-thick Ag gate electrode was deposited on top of the parylene film. Some devices have been annealed on a hot plate at 150 °C for 10 hours in air. The annealing, as it will be discussed below, improves the room-temperature mobility and significantly affects the temperature dependence of the mobility.

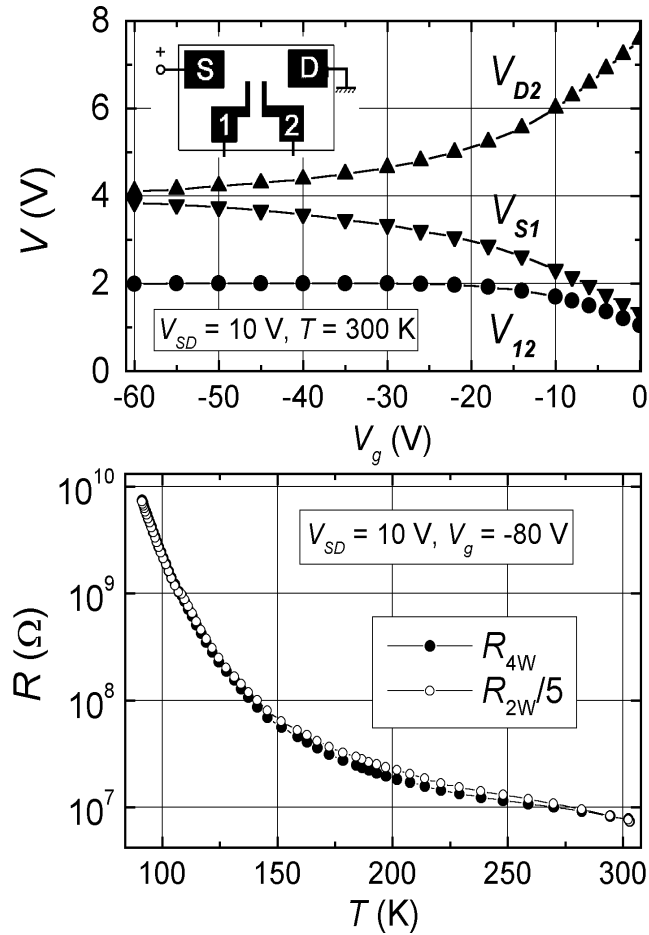

FIG. 2. The upper panel: voltage distribution in a 4-terminal OFET as a function of the gate voltage, measured at $V_{SD}$ = 10 V and $T$ = 300 K. Configuration of the contacts on the surface of the crystal, with two additional voltage probes 1 and 2, is shown schematically in the inset (the gate electrode is not shown). The lower panel: the temperature dependences of the 4-terminal resistance ($R_{4W}$) and the 2-terminal resistance divided by a factor of 5 (the distance between contacts 1 and 2 is approximately ~1/5 of the total channel length).

The room-temperature characteristics of a typical device are shown in Fig. 1. The source-drain distance $L$ is ~ 0.5 mm, the width of the device, $W$, is limited by the width of the crystal (~ 1 mm), the parylene thickness is ~ 0.2 μm. Sharp increase of the source-drain current, $I_{SD}$, at negative values of the gate voltage, $V_g$, indicates formation of the p-type conducting channel (the top panel). An important feature of this trans-conductance characteristic is a very small value of the switch-on voltage, $V_{SO}$ (the voltage of a sharp increase of $I_{SD}$) [10]. This almost threshold-less operation is due to high purity of the rubrene crystals [8], low concentration of defects and traps at the rubrene-parylene interface and good hole injection ability of the contacts. The $I_{SD}(V_{SD})$ characteristics at fixed $V_g$ (the lower panel of Fig. 1) exhibit a linear behavior at low $V_{SD}$ voltage and saturation of $I_{SD}$ at $|V_{SD}| \approx |V_g|$. These characteristics can be well described by the conventional FET equations [11].

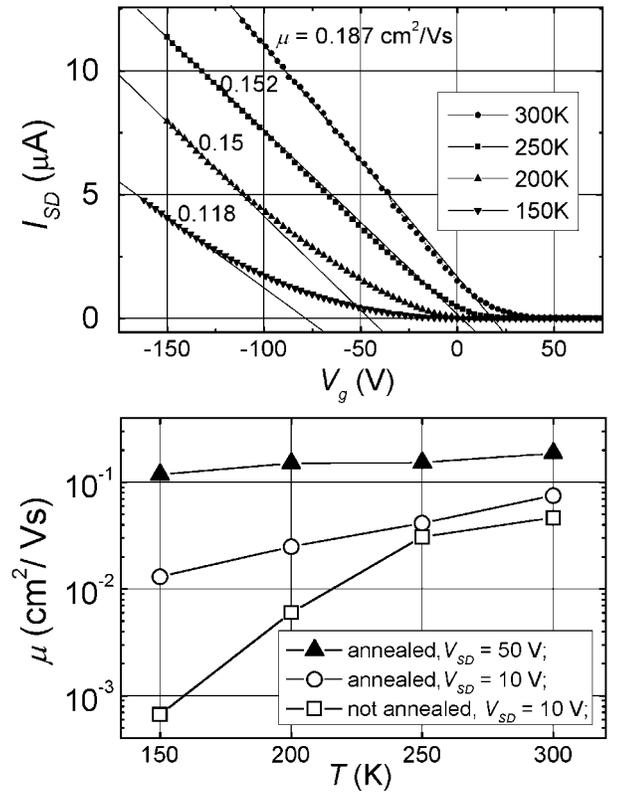

FIG. 3. The upper panel: $I_{SD}(V_g)$ characteristics of the OFET with evaporated silver source and drain contacts, measured at different temperatures for a fixed $V_{SD}$ = 50 V. The lower panel: $\mu(T)$ dependences, measured before annealing at $V_{SD}$ = 10 V (o) and after annealing at $V_{SD}$ = 10 V (O) and at 50 V (▲).

The data on Fig. 1 have been obtained in the two-contact measurements. Extraction of the field-induced carrier parameters from these data is possible if the resistance of the source and drain contacts is much smaller than the resistance of the conducting channel [12]. To verify this, we compared the results of two-contact and four-contact measurements.

Schematically, the four-contact configuration is shown in the inset to the upper panel of Fig. 2. The voltage drop on each pair of contacts was measured by the Keithley 6517 electrometer with the $10^{14}$ Ω input resistance as a function of the gate voltage. At $V_g$ = 0, almost all the source-drain voltage ($V_{SD} = 10$ V) is applied to the Schottky barrier between the drain electrode and the channel (the upper panel of Fig. 2). Increase of the carrier concentration with decreasing $V_g$ reduces the Schottky barrier resistance. At $V_g < -20$ V, the voltage $V_{12}$ saturates at 2 V. This indicates that the resistance of the source and drain contacts can be made negligible compared to the channel resistance, if a sufficiently large negative gate voltage is applied. Similar conclusion can be drawn from the comparison of the 4-terminal resistance ($R_{4W} \equiv V_{12}/I_{SD}$) with the 2-terminal resistance ($R_{2W} \equiv V_{SD}/I_{SD}$). The temperature dependences of $R_{4W}$ and $R_{2W}$ measured over the range $T = 77 - 300$ K at $V_g = -80$ V are identical (the lower panel of Fig. 2), the ratio $R_{4W}/R_{2W}$ is consistent with the channel geometry. The data discussed below have been obtained in the regimes when the contact resistance can be neglected.

Evolution of the trans-conductance dependences $I_{SD}(V_g)$ with temperature for a device with thin-film silver contacts is shown in Fig. 3 (the upper panel). In these measurements, $V_{SD} = 50$ V was chosen to ensure a small contact resistance. Because of a large value of $V_{SD}$, the pinch-off of the conducting channel at room temperature occurs at a large positive gate voltage [13]. The values of $V_g$, obtained by extrapolation of the linear portion of the $I_{SD}(V_g)$ to $I_{SD} = 0$ (the upper panel in Fig. 3), decrease with cooling and eventually become negative. This suggests an increase of the threshold voltage with cooling due to depopulation of the carrier traps at the rubrene-parylene interface with decreasing the temperature.

The field-effect mobility, $\mu$, can be estimated from the $I_{SD}(V_g)$ dependence as:

$$\mu = (L/WC_iV_{SD})(dI_{SD}/dV_g), \qquad (1)$$

where $C_i$ is the capacitance per unit area of the gate insulator. The mobility typically increases with $V_{SD}$ at small bias voltages and becomes $V_{SD}$-independent for the 0.5 – 1 mm-long channels at $V_{SD} > 10 – 30$ V. At sufficiently large negative $V_g$, the dependences $I_{SD}(V_g)$ are linear (Fig. 3, the upper panel), and the mobility is $V_g$-independent. The mobility decreases with cooling: the temperature dependences of $\mu$, measured at different values of $V_{SD}$, are shown in the lower panel of Fig. 3.

It has been observed that the post-fabrication annealing improves the mobility, especially at low temperatures (the lower panel of Fig. 3). This fact allows us to trace evolution of $\mu(T)$ for the same device as a function of its room temperature mobility $\mu(300$ K). Small increase of $\mu(300$ K) results in a substantial increase of the low temperature mobility, which is consistent with the results reported for the thin-film OFETs [4]. Thus, there is a possibility that with a further increase of $\mu(300$ K), the crossover from the thermally activated transport to a band-like electrical conduction could be observed at low temperatures.

In conclusion, by using parylene as a material for the gate insulator, we successfully fabricated the field-effect transistors at the surface of single crystals of rubrene. The devices demonstrate the hole-type conductivity with the mobility up to 1 cm$^2$/Vs and the on/off ratio up to $10^4$ at room temperature. The temperature dependence of the mobility depends strongly on the value of $\mu(300$ K). Further improvement of the quality of the rubrene crystals and the rubrene-parylene interface may lead to the observation of the increasing mobility with decreasing temperature.

We thank Ch. Kloc for helpful discussions. This work was supported in part by the NSF grant DMR-0077825, the DOD MURI grant DAAD19-99-1-0215, and the NATO Collaborative Linkage grant PST.CLG.979275.